\documentclass[%
reprint,
 amsmath,amssymb,
 aps,
prb,
]{revtex4-1}

\usepackage{graphicx}
\usepackage{dcolumn}
\usepackage{bm}
\usepackage{placeins}
\usepackage{xcolor}
\usepackage{tabularx}
\usepackage{multirow}

\begin{document}

\title{Lattice dynamics of palladium
in the presence of electronic correlations}
\author{W.~H.~Appelt$^{a,b}$, A.~\"Ostlin$^{c}$, I.~Di~Marco$^{d,e,f}$,  I.~Leonov$^{g,h}$,
M.~Sekania$^{c,i}$, D.~Vollhardt$^{c}$, L.~Chioncel$^{b,c}$}

\affiliation{$^{a}$Theoretical Physics II, Institute of Physics, University of
Augsburg, D-86135 Augsburg, Germany}
\affiliation{$^{b}$Augsburg Center for Innovative Technologies, University of Augsburg,
D-86135 Augsburg, Germany}
\affiliation{$^{c}$Theoretical Physics III, Center for Electronic
Correlations and Magnetism, Institute of Physics, University of
Augsburg, D-86135 Augsburg, Germany}
\affiliation{$^{d}$ Asia Pacific Center for Theoretical Physics, Pohang, Gyeonbuk 790-784, Republic of Korea}
\affiliation{$^{e}$ Department of Physics, POSTECH, Pohang, Gyeonbuk 790-784, Republic of Korea}
\affiliation{$^{f}$ Department of Physics and Astronomy, Uppsala University, Box 516, SE-75120 Uppsala, Sweden}
\affiliation{$^g$
Institute of Metal Physics, Sofia Kovalevskaya Street 18, 620219 Yekaterinburg GSP-170, Russia}
\affiliation{$^h$
Materials Modeling and Development Laboratory, NUST "MISiS", 119049 Moscow, Russia}
\affiliation{$^i$ Andronikashvili Institute of Physics, Tamarashvili 6, 0177 Tbilisi, Georgia}

\date{\today}

\begin{abstract}
We compute the phonon dispersion, density of states, and the Gr\"uneisen
parameters of bulk palladium in the combined density functional theory (DFT) and
dynamical mean-field theory (DMFT).
We find good agreement with experimental results for ground state properties (equilibrium lattice
parameter and bulk modulus) and the experimentally measured phonon
spectra.
We demonstrate that at temperatures $T \lesssim 20~K$ the
phonon frequency in the vicinity of the Kohn anomaly,
$\omega_{T1}({\bf q}_{K})$,  strongly decreases. This is in contrast to DFT where this frequency remains essentially constant in the whole temperature range.
Apparently correlation effects
reduce the restoring force of the ionic displacements at low temperatures,
leading to a mode softening.
\end{abstract}

\maketitle

\section{\label{sec:intro}Introduction}

The vibrational spectrum is a key quantity for understanding the interatomic
interactions in a material. For solids, various inelastic scattering techniques
exist to measure the phonon dispersion relation. These results can be readily
compared with calculations within density functional theory (DFT)~\cite{ho.ko.64,kohn.99,jo.gu.89,jone.15}.
Within the DFT, a frequently used method to compute the phonon dispersion relation
is the frozen phonon approach based on total energy calculations. The same
formalism is also used to compute ground state properties such as
equilibrium lattice parameters and bulk modulus. In the last decade it has
been demonstrated that within the combined DFT and dynamical mean-field
theory (DMFT)~\cite{me.vo.89,go.ko.92,ge.ko.96,ko.vo.04}, the so called DFT+DMFT approach~\cite{held.07,ko.sa.06}, many of the electronic ground state
properties of $d$-metal elements, their alloys and compounds can be
properly described~\cite{ko.sa.06,held.07,ka.ir.08,ku.le.17}.
This is also true in the case of palladium where results of the local density approximation (LDA) for the bulk
modulus turns out to overestimate the experimental value by more than $10\%$. Indeed, using LDA+DMFT
it was shown~\cite{oe.ap.16} that the experimental volume $V=99.3\,$a.u.$^3$
is reproduced and that the difference in the bulk modulus from the
experimental value is reduced to less than $1\,\%$. In addition,
the results for the spectral function demonstrate the existence of a
high-energy satellite formation in agreement with  experiment~\cite{ch.81}.
In our previous study we also discussed the importance of local and
nonlocal correlation effects~\cite{oe.ap.16} by comparing the LDA+DMFT results
with the self-consistent quasiparticle GW approach~\cite{sc.06,ko.07}.
More recently, a LDA+DMFT study~\cite{sa.re.18} using a quantum Monte Carlo impurity solver found
comparable electronic effective mass as in Ref.~\onlinecite{oe.ap.16}.
In addition in Ref.~\onlinecite{sa.re.18} employing a lattice (${\bf k}$-dependent) FLEX solver
resulted in a self-energy with weak ${\bf k}$-dependence.
These results show the importance of including local, dynamic correlation
effects in computing the ground state
properties of Pd.

The goal of this paper is to calculate the changes in the phonon dispersions
and phonon density of states of elemental solid palladium under pressure when electronic correlation effects are included.
In addition, we investigate the volume dependence of the phonon frequencies, i.e. the mode Gr\"uneisen parameters.
We employ the quasi-harmonic approximation (QHA), according to which the harmonic approximation holds at every volume, and
the harmonic frequencies are replaced by renormalized volume dependent
frequencies~\cite{do.93,ho.74}.
It has been shown in the case of Pd that the QHA describes
thermal dilation effects accurately and that higher order corrections are
three orders of magnitude smaller for temperatures up to the melting
temperature~\cite{zo2.90}.

Early experimental studies of the phonon spectra in palladium were performed by Miiller and
Brockhouse~\cite{mii.bro.68,mii.bro.71,mi.75} using neutron scattering,
and a Kohn anomaly
at a vector around ${\mathbf q} = \frac{2 \pi}{a}[0.35,0.35,0]$ was proposed.
The analysis by Freeman~\textit{et~al.}~\cite{fr.wa.79} showed a pronounced
enhancement of the
generalized susceptibility (Lindhard function) at a $\mathbf{q}$-vector of
around $\frac{2\pi}{a}[0.325,0.325,0]$.
Later, Savrasov and Savrasov~\cite{sa.sa.96} and Takezawa~\cite{ta.na.05}
reported numerical studies where they found no signature of phonon anomalies in Pd.
Stewart~\cite{st.08} reported a softening in the phonon dispersion at around
${\mathbf q}=\frac{2\pi}{a}[0.3,0.3,0]$ using DFT, which still somewhat
underestimated the experimental value. He showed that the electronic degrees
of freedom are responsible for the softening of the phonon mode.
However, Liu \textit{et~al.}~\cite{li.ya.11} pointed out that the observed
Kohn anomaly by Stewart depends sensitively on the technical details of the
simulation. They observed that the Kohn anomaly vanishes when a
calculation is performed with a denser $\mathbf{k}$-mesh.
Recently, Dal Corso \cite{da.13} investigated the influence of various
exchange correlation functionals on the phonon spectrum in transition and noble
metals. This included the zero-point anharmonic expansion
(ZPAE)~\footnote{see for example chapter 7 Ref.~\onlinecite{gr.99}} and the thermal expansion of the lattice constant, which also affects the equilibrium bulk modulus~\cite{gr.hi.07,st.04}.
He showed that the LDA calculations overestimate the experimentally measured phonon frequencies by $5\,\%$ while the generalized gradient approximation (GGA)~\cite{pbe1} underestimates them by about $7\,\%$ at the high symmetry point $L=\frac{\pi}{a}[1,1,1]$.

Until now, the numerical modeling of phonons in Pd did not include correlation
effects. Following up our previous work, in this article we compare
the results of the lattice dynamics obtained by LDA and LDA+DMFT.
In Sec.~\ref{sec:comp_det} we discuss the details of the calculation
and in Sec.~\ref{sec:froz_phon} we briefly describe the methodology of frozen phonon calculations within the LDA and LDA+DMFT method.
In Sec.~\ref{subsec:ad} the phonon dispersion curves are computed within the Born-Oppenheimer (BO) adiabatic approximation including DMFT corrections for the electrons by assuming the system to be perfectly harmonic.
In Sec.~\ref{subsec:gr} we calculate the mode Gr\"uneisen parameters which can be employed to assess the anharmonicity on the level of the QHA. Namely, although the QHA neglects cubic or quartic terms in the expansion of the Born-Oppenheimer energy surface, it includes the anharmonic strain- or volume-dependence of the frequencies.
Sec.~\ref{sec:Tdependence} addresses the finite
temperature behaviour of a specific mode close to the Kohn
anomaly. The phonon frequencies computed within LDA+DMFT show a strong temperature dependence which is not observed in DFT(LDA).
The article is closed with a brief conclusion in Sec.~\ref{sec:conc}.

\section{\label{sec:comp_det}Computational Details}

Pd crystallizes in the face-centered cubic structure ($Fm\overline{3}m$ space group) with a one-atom basis.
The palladium atom is located at the Wyckoff position $4a$, i.e. $(0,0,0)$.
We performed a lattice relaxation within
LDA with the parametrization of Perdew and Wang~\cite{pe.wa.92} for the exchange-correlation functional.
In the present study we employ the
full-potential linearized muffin-tin orbitals (FPLMTO) method, as implemented
in the {\sc RSPt} code~\cite{rsptbook,gr.ma.12,di.mi.09}.
We used three kinetic energy tails with corresponding energies $-0.3$, $-2.3$, and
$-1.5\,$Ry.
The  muffin-tin radius was set to $2.45\,$a.u. and was kept constant for all unit-cell volumes.
For the charge density and the angular decomposition of the potential inside the muffin-tin spheres a maximum angular
momentum $l_\text{max} = 8$ was set.
The calculations include spin-orbit coupling and scalar-relativistic terms.
Brillouin-zone integration was performed using a Fermi-Dirac function (thermal broadening).
The ${\mathbf{k}}$-mesh grid is $16 \times 16 \times 16$ for the equations of state.
To go beyond the LDA and include electronic correlations
the local interaction term
$1/2\sum_{\{\lambda_i\}}U_{\lambda_1\lambda_2\lambda_4\lambda_3}
 c^\dagger_{\lambda_1}c^\dagger_{\lambda_2}c_{\lambda_3}c_{\lambda_4} $
is added to the Hamiltonian, where $\lambda_i=(m_i,\sigma_i)$.
Here the  operators $c^{(\dagger)}_{\lambda_i}$ annihilate (create) electrons on the
orbital $m_i$ with the spin $z$-projection $\sigma_i=\uparrow,\downarrow$.
The summation over the set of basis functions $\{m_i=-l,\dots,l\}$  ($l=2$) is restricted to the correlated subset (the $4d$-orbital subspace in our case).
We used the so-called muffin-tin-only correlated subspace which corresponds to basis functions which are zero outside the muffin tin region;
see Sec. II in Ref.~\onlinecite{gr.ma.07} and Ref.~\onlinecite{gr.ma.12} for additional information about the basis-function dependence.
The interaction matrix elements $U_{\lambda_1\lambda_2\lambda_4\lambda_3}$ are
components of a rotational invariant fourth-rank tensor.
These are computed from the on-site Slater-Koster integrals using atomic wave functions~\cite{co.du.16}.
The averaged local Hubbard interaction $U$ is varied in the range from $0.75\,$eV to $1.30$~eV, while the Hund's rule coupling is kept fixed at $J=0.3\,$eV.
A double-counting correction is included in order to cancel the contributions due to the total energy functional originating from the electron-electron interaction captured within the exchange-correlation functional of DFT.
This correction is applied to the self-energy~\cite{gr.ma.12}:
$\Sigma_{\lambda_1\lambda_2}(i\omega_n)\to \Sigma_{\lambda_1\lambda_2}(i\omega_n)-\Sigma_{\lambda_1\lambda_2}^{DC}(0)$ with
$\Sigma_{\lambda_1\lambda_2}^{DC}(0)=\delta_{m_1m_2}/(2l+1)\sum_{m_3} \Sigma_{m_3\sigma_1,m_3\sigma_2}(0)$, since this is known to be the appropriate correction for metals, which ensures Fermi-liquid behavior~\cite{pe.ma.03}.
We note that the frozen-phonon approach is not limited to a specific choice
of double-counting.
The Spin-polarized T-matrix Fluctuation
Exchange Approximation (Sp-TFLEX)~\cite{li.ka.97} on the Matsubara domain
was used as an impurity solver for the DMFT problem.
Throughout this paper finite temperatures are only
considered for the electronic subsystem, where the temperature enters in the Matsubara frequencies
$\omega_n=(2n+1)\pi T$, ($n\in \mathcal{Z}$).
We chose the same temperature for the imaginary-frequency Matsubara mesh in DMFT as for the Brillouin-zone integration.
The electronic charge was updated self-consistently in all LDA and LDA+DMFT calculations~\cite{gr.ma.12}.
The LDA+DMFT results do not change significantly in the
temperature range from 275 to 325 K, which allows for a comparison between
our results ($T=316\,$K) and the experimental data measured at room
temperature.
In the search of a mode softening
in the vicinity of the Kohn anomaly we also performed a finite temperature
study in an extended temperature range down to $15\,$K.
%

For the frozen phonon superlattice we used the Brillouin-zone integration by Fourier quadrature, which permits us to keep the $\mathbf{k}$-mesh density (namely $16 \times 16 \times 16$ $\mathbf{k}$-points per fcc Brillouin zone) consistent in all calculations~\cite{fr.89}.
This minimizes systematic numerical errors when total energies are compared.
We estimated errors due to ${\mathbf{k}}$-mesh sampling for the LDA study by computing all phonon frequencies using a denser ${\mathbf{k}}$-mesh of $32 \times 32 \times 32$ within the fcc Brillouin-zone for the frozen phonon calculations.
The error estimates for the phonon frequencies were computed as the differences $\Delta_\nu(\mathbf{q})=|\omega_\nu(\mathbf{q})_\text{fine}-\omega_\nu(\mathbf{q})_\text{sparse}|$, where $\omega_\nu(\mathbf{q})_{\text{fine}}$ and $\omega_\nu(\mathbf{q})_{\text{sparse}}$ denote the phonon frequencies which are computed using the $32 \times 32 \times 32$ and $16 \times 16 \times 16$ ${\mathbf{k}}$-mesh, respectively.
We found that $\Delta_\nu(\mathbf{q})$ is smaller than  $3\,$meV.
In the presented Figures, we indicate the errors due to finite \textbf{k}-mesh sampling.
The same error estimates have been performed for the LDA+DMFT calculations
for the zone-boundary phonon frequencies.
The LDA+DMFT error estimates are found to be smaller than their corresponding LDA values.
The reduction of the sensitivity of
the LDA+DMFT results as compared to the LDA results is expected to originate
from the broadening of the one-electron spectra due to many-body
effects.
A fine ${\mathbf{q}}$-mesh grid of $24 \times 24 \times 24$  is used to compute the phonon density of states (DOS) employing the tetrahedron integration method~\cite{la.vi.84} after fitting to a five nearest-neighbor
Born-von Karman (BvK) force constant model.

\section{\label{sec:froz_phon}Phonon calculations from first principles}
The frozen phonon approach~\cite{ku.81,yi.82} is frequently used
to calculate the phonon spectra.
In this method the eigenvector of the given phonon is obtained from symmetry considerations when the
appropriate atomic displacements are frozen in. The associated lowering of the
symmetry corresponds to supercells of dimensions that match  the
phonon wavelength.
In this way the method is not limited to the linear response of the energy
functional with respect to lattice vibrations.
A shortcoming of the frozen-phonon method is the supercell size, which may become
impractically large, especially when long-wavelength phonons are considered.

The linear response method in combination with LDA+DMFT was first applied on the Mott insulators, NiO and MnO~\cite{sa.ko.03}, and
is currently being further extended~\cite{le.an.14,ha.pa.16,ha.18}.
In combination with the many-body (LDA/GGA)+DMFT approach the frozen
phonon technique was used to compute the phonon spectrum of the elemental
metals, namely Fe (Ref.~\onlinecite{le.po.12}) and Pu (Ref.~\onlinecite{da.sa.03}),
as well as the compound KCuF$_3$ (Ref.~\onlinecite{le.bi.08}).
Below we take the frozen-phonon approach and present results for the
phonon band structure and the corresponding phonon density
of states of palladium using the LDA and LDA+DMFT approach.

\begin{figure}[ht]
\includegraphics[width=\columnwidth,clip=true]{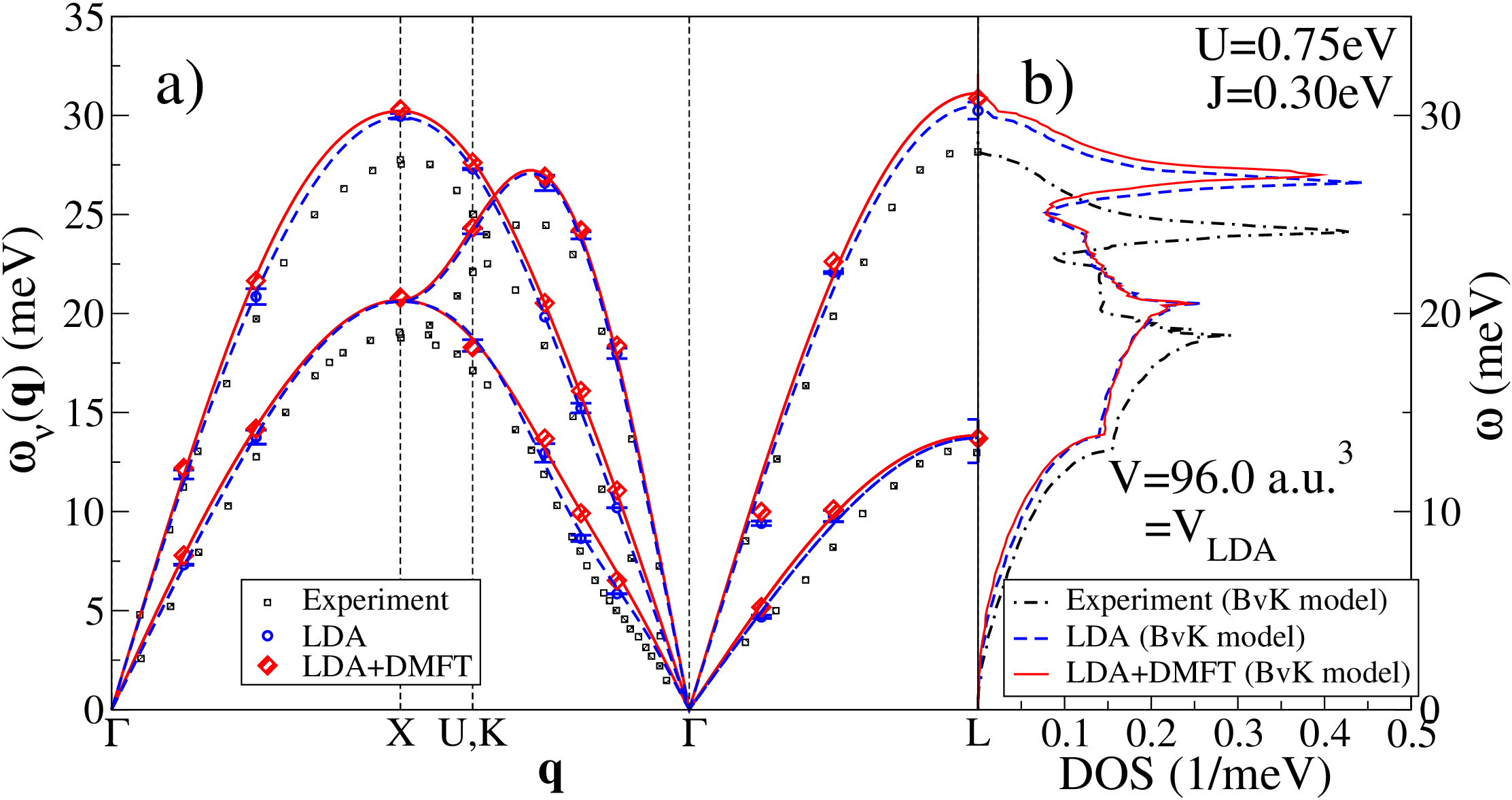} \\
\includegraphics[width=\columnwidth,clip=true]{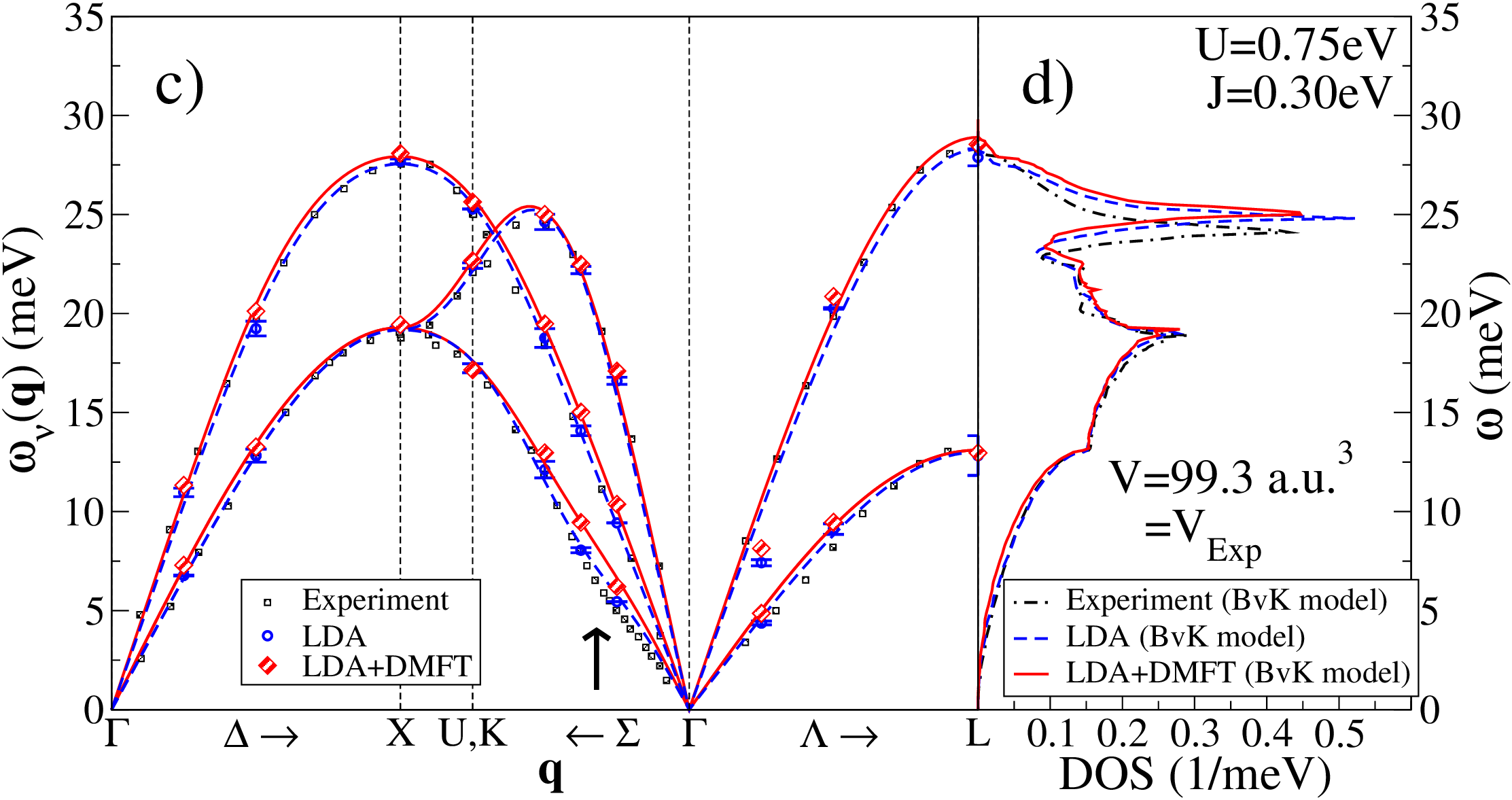}
\caption{Computed phonon dispersions along the main symmetry directions
for the LDA equilibrium volume (Fig.~\ref{Fig1}a) and the
experimental volume (Fig.~\ref{Fig1}c).
The computed phonon spectra in LDA and LDA+DMFT are shown with blue and red symbols
respectively.
The experimental dispersion curves of Pd measured at room temperature ($296\,$K) are taken from Ref.~\onlinecite{mii.bro.68} (black boxes).
The solid lines are fits to a five nearest-neighbor Born-von Karman (BvK) force constant model.
The Fermi-surface nesting vector $\mathbf{q}_\text{nest}=\frac{2\pi}{a}[0.32,0.32,0]$ is indicated by the vertical arrow.
Figs.~\ref{Fig1}b and ~\ref{Fig1}d: Computed phonon DOS at the LDA equilibrium volume and experimental equilibrium volume, respectively.
        }
        \label{Fig1}
\end{figure}

\subsection{Phonon dispersions and density of states}
\label{subsec:ad}
In the BO approximation the nuclei are considered immobile during the characteristic electronic time scales.
Within the harmonic approximation the potential energy surface is expanded up to second order in the ionic displacement.
The equation of motion for the ionic subsystem is determined by the BO energy surface, obtained from the total energy within the LDA and LDA+DMFT.
Its solution specifies the eigenmodes of a vibration with
the characteristic frequencies $\omega_{\nu}(\mathbf{q})$, where the index
$\nu=1,2,3$ enumerates the three linear independent phonon branches. The relation between the symmetry adapted basis $|\nu\rangle$ and the Cartesian basis is explained in Appendix~\ref{app:symmetry}.

\begin{table*}
\caption{ Left part: Volume $V$ for various methods (upper part) and experiments (lower part). Middle part: Computed or experimentally measured pressure $P$, and isothermal/adiabatic bulk modulus $K_{x=T/S}$ in the upper part and lower part, respectively.
The computed values for the bulk modulus $K_x$ includes a correction term for the pressurized sample. The computations where done at $T=316\,$K and the experiments have been performed at room temperature as well.
Right part: Zone boundary frequencies computed under adiabatic conditions are compared with the experimental phonon frequencies~\cite{mii.bro.68}.
}
  \begin{ruledtabular}
    \begin{tabular}{l||r||r|r|c|l||rrrr}
\multirow{2}{*}{method}  & \multirow{2}{*}{$V\,$(a.u.$^3$)} & \multirow{2}{*}{$P$(GPa)} &  \multirow{2}{*}{$K_x\,$(GPa)$^a$}& \multirow{2}{*}{$x$}  & \multirow{2}{*}{Refs}  & \multicolumn{4}{c}{$\omega_\nu(\mathbf{q})$} \\ \cline{7-10}
  & &  & & & & $(L,L_2^{-})$ & $(L,L_3^{-})$ & $(X,X_3^{-})$ & $(X,X_5^{-})$   \\  \hline \hline
  LDA  & 95.94& 0 &  226.6 & T/S & [\onlinecite{oe.ap.16}]  & 30.2 & 13.6  &  30.0 & 20.7 \\
   LDA   & 99.30& -10.3& 186.5 & T/S & [\onlinecite{oe.ap.16}]  & 27.9 & 12.8 & 27.7 & 19.3  \\
  LDA+DMFT  & 95.94 & 4.5 & 225.7 & S & [\onlinecite{oe.ap.16}]  &  30.9 & 13.7 &  30.3 & 20.8 \\
  LDA+DMFT  & 99.30& -6.0 & 186.0  & S & [\onlinecite{oe.ap.16}] & 28.5 & 13.0 & 28.1 & 19.4 \\
  \hline
     Exp   &  99.30& 0& 195 & S & [\onlinecite{mii.bro.68},\onlinecite{si.79}] $^{b}$  & 28.2 & 13.0 & 27.5 & 18.8\\
  Exp    &  99.30&0& 189 & T & [\onlinecite{gs.64}]   & - & - & - & - \\
    Exp    & 99.30& 0 & 193 &T  & [\onlinecite{so.93}] & - & - &  - & - \\
\end{tabular}
  \end{ruledtabular}
  \begin{flushleft}
   $^a$ The values in the $K_x$ column correspond either to the adiabatic ($K_{x=S}$) value, the isothermal ($K_{x=T}$) value, or to both when $K_T$ and $K_S$ coincide in the $T=0$ formalism. The difference between the reciprocal values $1/K_S$ (Ref.~\onlinecite{mii.bro.68}) and $1/K_T$ (Ref.~\onlinecite{gs.64}) is $1/K_T-1/K_S=VT\beta^2/C_P$, where $C_P$ is the heat capacity and constant pressure, and $\beta=1/V\left(\partial V/\partial T \right)_P$ is the volumetric thermal expansion coefficient.
   $^b$ The bulk modulus is computed from the elastic constants (see Table I in Ref.~\onlinecite{si.79}).
     \end{flushleft}
   \label{tab:elast}
\end{table*}

In Figs.~\ref{Fig1}a and~\ref{Fig1}c we present
the phonon dispersion curves computed at the LDA equilibrium volume and at the experimental equilibrium volume, respectively, in comparison with the experimental measurements~\cite{mii.bro.68}.
We find an good qualitative agreement with the experimentally measured dispersion curves (see also Table~\ref{tab:elast}).
For the LDA equilibrium volume ($V_{\rm LDA}$) the phonon frequencies are overestimated
for most of the ${\mathbf q}$-points.
This overestimation can be corrected considering the experimental equilibrium
volume $V_{\rm Exp}$ which is larger than $V_{\rm LDA}$~\cite{oe.ap.16}.
Hence the phonon frequencies computed with $V_{\rm Exp}$ correspond to the values which would be obtained for a negative pressure of $P=-10.3\,$GPa (see also Table~\ref{tab:elast}).
We find that by including electronic interactions of $U=0.75\,$eV and $J=0.3\,$eV does not improve the LDA results further.
The bulk modulus and lattice constant computed within LDA+DMFT, however, show a better agreement with experiment than the LDA result~\cite{oe.ap.16}.
One significant difference is observed along the $[110]$-direction (see Figs.~\ref{Fig1}a and~\ref{Fig1}c), namely
the mode softening obtained within the LDA at ${\bf q}=\frac{2\pi}{a}[0.32,0.32,0]$ in the $T_1$ branch (see Appendix~\ref{app:symmetry} for definition of the branch labeling) is reduced within LDA+DMFT.
Hence the presented result, computed at high temperatures, is not in line with the experiment.
A detailed study on the temperature dependence of the mode
in the vicinity of ${\bf q}=\frac{2\pi}{a}[0.32,0.32,0]$
will be presented in Sec.~\ref{sec:Tdependence}. There it is shown that at lower temperatures ($T\leq15\,$K) the mode softening is well described within LDA+DMFT.

In Figs.~\ref{Fig1}b and~\ref{Fig1}d we present
the phonon DOS computed at the LDA equilibrium volume and at the experimental equilibrium volume, respectively, and compare the LDA and LDA+DMFT phonon DOS with the experimental results.
The main features of the experimental DOS are the two peaks at around 20 and 25 meV.
At $V=96.0\,$a.u.$^3$ we find that both within the LDA and LDA+DMFT  the peak positions are shifted up compared to the experimental phonon DOS.
At $V=99.3\,$a.u.$^3$ the lower peak (around 20 meV) is captured by both the LDA and LDA+DMFT method, while the higher peak (around 25 meV) is shifted up in both methods.
In LDA+DMFT the peak position (around 25 meV) in the DOS agrees well with the experiment, since it is only shifted by about $1\,$meV.

\begin{figure*}[ht]
\includegraphics[width=\columnwidth,clip=true]{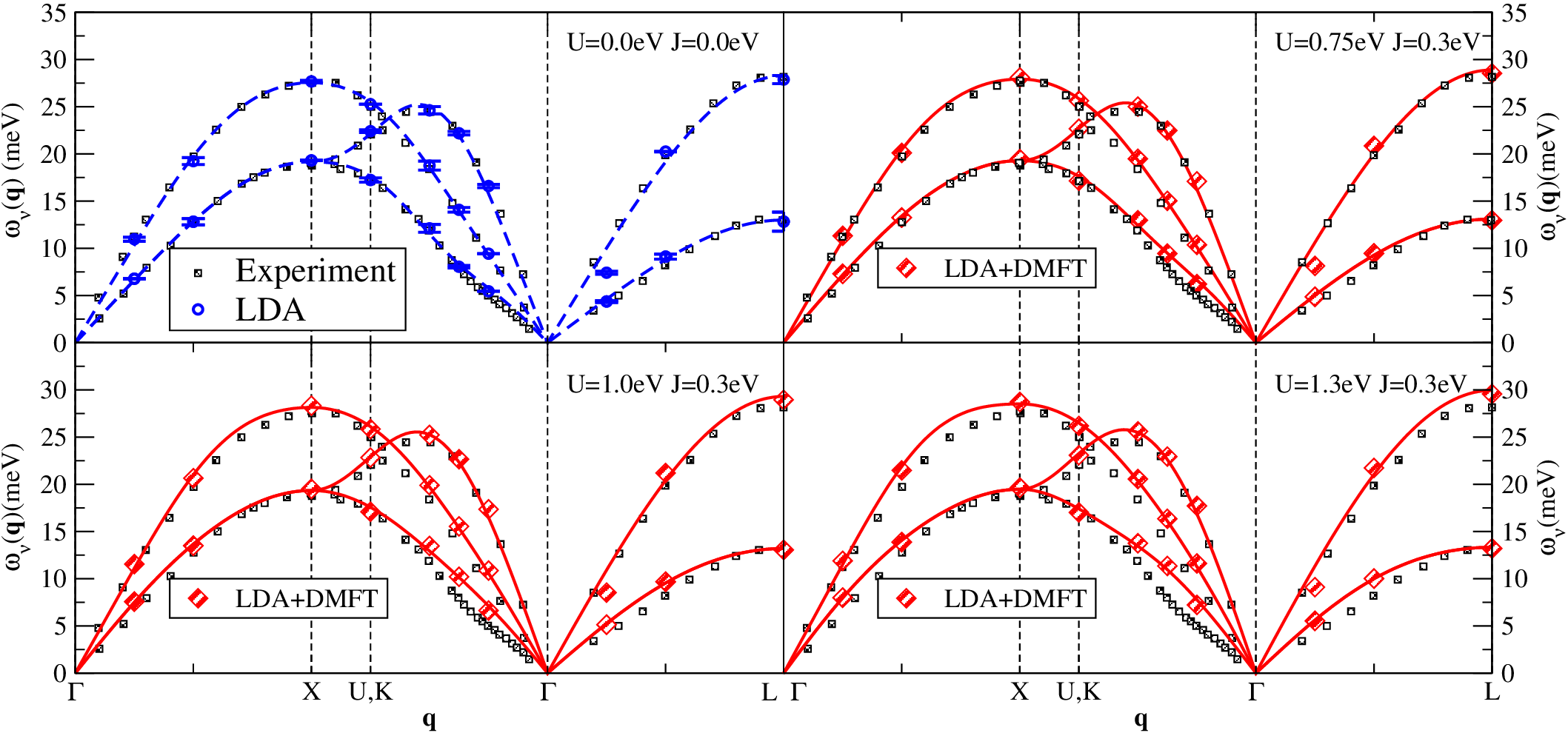}
\caption{Computed phonon dispersion curves along the main symmetry directions for the experimental volume for different values of $U$ and $J$, where the LDA result corresponds to the case when $U=J=0\,$eV.
The experimental dispersion curves of Pd recorded at room temperature are taken from Ref.~\onlinecite{mii.bro.68} (black boxes).
The connecting lines were obtained by fitting the data to a five nearest-neighbor Born-von Karman force constant model.
        }
        \label{Fig2}
\end{figure*}

In Fig.~\ref{Fig2} we compare phonon spectra computed at $316\,$K for different values of the local Hubbard interaction $U$  with the experimental results measured at room temperature \cite{mii.bro.68}.
For $U=0.75\,$eV and $J=0.3\,$eV the overall agreement with the experimental phonon dispersion is good, differences being similar to those in the case of the LDA results.
For $U=1.0\,$eV and $U=1.3\,$eV at $T=316$~K  we find that the phonon frequencies are higher than those obtained within the LDA. Furthermore, the phonon frequencies are found to increase with increasing  Coulomb interaction strength, i.e., the modes become stiffer.

The softening of the phonon dispersion along the $\Sigma$-line ($\mathbf{q}_\text{nest}=\frac{2\pi}{a}[0.32,0.32,0]$) is not described by the  LDA+DMFT results for any of the investigated interaction strengths.
This will be further discussed in Sec.~\ref{sec:Tdependence}, where we address the finite temperature behavior of the phonon frequencies.

\subsection{Phonon dispersion at different volumes}
\label{subsec:gr}

\begin{table*}
\caption{Left part: Gr\"uneisen parameter $\gamma_\nu^G({\mathbf q})$ computed in LDA and LDA+DMFT, the  difference between frequencies calculated in LDA and LDA+DMFT $\delta \omega_\nu(\mathbf{q})=\omega_\nu(\mathbf{q})^{\mathrm{DMFT}}-\omega_\nu(\mathbf{q})^{\mathrm{LDA}}$ computed for the branches $\nu=L_2^-,L_3^-,X_3^-,X_5^-$. Right part: Average Gr\"uneisen parameter $\langle \gamma^G\rangle$, and thermodynamic Gr\"uneisen parameter $\gamma_{td}^G$ computed in both methods.}
  \begin{ruledtabular}
  \begin{tabularx}{\textwidth}{ll|rrrr||rrr}
 method &  & $(L,L_2^{-})$ & $(L,L_3^{-})$& $(X,X_3^{-})$& $(X,X_5^{-})$& $\langle \gamma^G({\mathbf q})\rangle$ &  $\gamma_{td}^G$ ($T=298\,$K) &   $\gamma_{td}^G$ ($T=316\,$K) \\
  \hline
LDA: &$\gamma_\nu^G(\mathbf{q})$ & $2.34$ & $1.68$ & $2.29$ & $2.01$ & 2.210 & 2.216 & 2.217 \\
LDA+DMFT: &$\gamma_\nu^G(\mathbf{q})$ & $2.28$ & $1.74$ & $2.24$ & $2.01$ & 2.019 & 2.020 & 2.020\\
 Difference & $\delta \omega_\nu(\mathbf{q})$ & $0.45$ & $-0.15$ & $0.39$ & $0.13$ &  & &\\
  \end{tabularx}
  \end{ruledtabular}
   \label{tab:gruen}
\end{table*}

In the first attempt to analyze the volume dependence of the phonon frequencies on a quantitative level, it is useful to consider the  Gr\"uneisen parameter $\gamma_\nu^G({\mathbf q})$.
It represents the relative change of $\omega_\nu(\mathbf{q})$ with isotropic change in volume for a particular phonon branch $\nu$ at the symmetry point ${\mathbf q}$, and is computed according to:
\begin{eqnarray}\label{eq:omega}
\gamma_\nu^G({\mathbf q})&=& -\frac{V}{\omega_\nu({\mathbf q})}\left(\frac{\partial \omega_\nu({\mathbf q})}{\partial  V}\right).
\end{eqnarray}
Usually the values of $\gamma_\nu^G({\mathbf q})$ are
positive and lie in the range $1.5\pm 1.0$ (Ref.~\onlinecite{gr.99}), where a larger value suggests that anharmonic effects are important for the particular phonon mode of interest. This is only valid when anharmonicity is weak to start with.
Using the values provided by Eq.~\ref{eq:omega}, we computed the average Gr\"uneisen parameter $\langle \gamma^G\rangle$ defined as $\langle \gamma^G\rangle=\sum_{\nu}1/3V_{BZ}\int d\mathbf{q} \gamma_\nu^G({\mathbf q})$.
We also computed the thermodynamic Gr\"uneisen parameter which
within the QHA takes the form: $\gamma_{td}^G=\sum_{\mathbf{q},\nu} C_V(\mathbf{q},\nu) \gamma^G_\nu (\mathbf{q})/\sum_{\mathbf{q},\nu} C_V(\mathbf{q},\nu)$.
The specific heat per mode has the form $C_V(\mathbf{q},\nu)=k_B x^2 e^x/(e^x-1)^2$, with $x=\hbar \beta\omega_\nu(\mathbf{q})$.
We find that $\gamma_{td}^G$ is essentially equivalent to the average value $\langle \gamma^G\rangle$ (see Tab.~\ref{tab:gruen}).
The main effect of electronic correlations is a decrease of $\gamma_{td}^G$ by about $10\%$.
We provide these results with the following interpretation: We know from previous studies\cite{oe.ap.16} that electronic correlations lead to a softening of the lattice it can be expected that the coupling between phonon degrees of freedom and the volume are reduced.
Hence the Gr\"uneisen parameter is expected to be reduced by correlation and that is what is seen.

\begin{figure}[t]
\includegraphics[width=\columnwidth,clip=true]{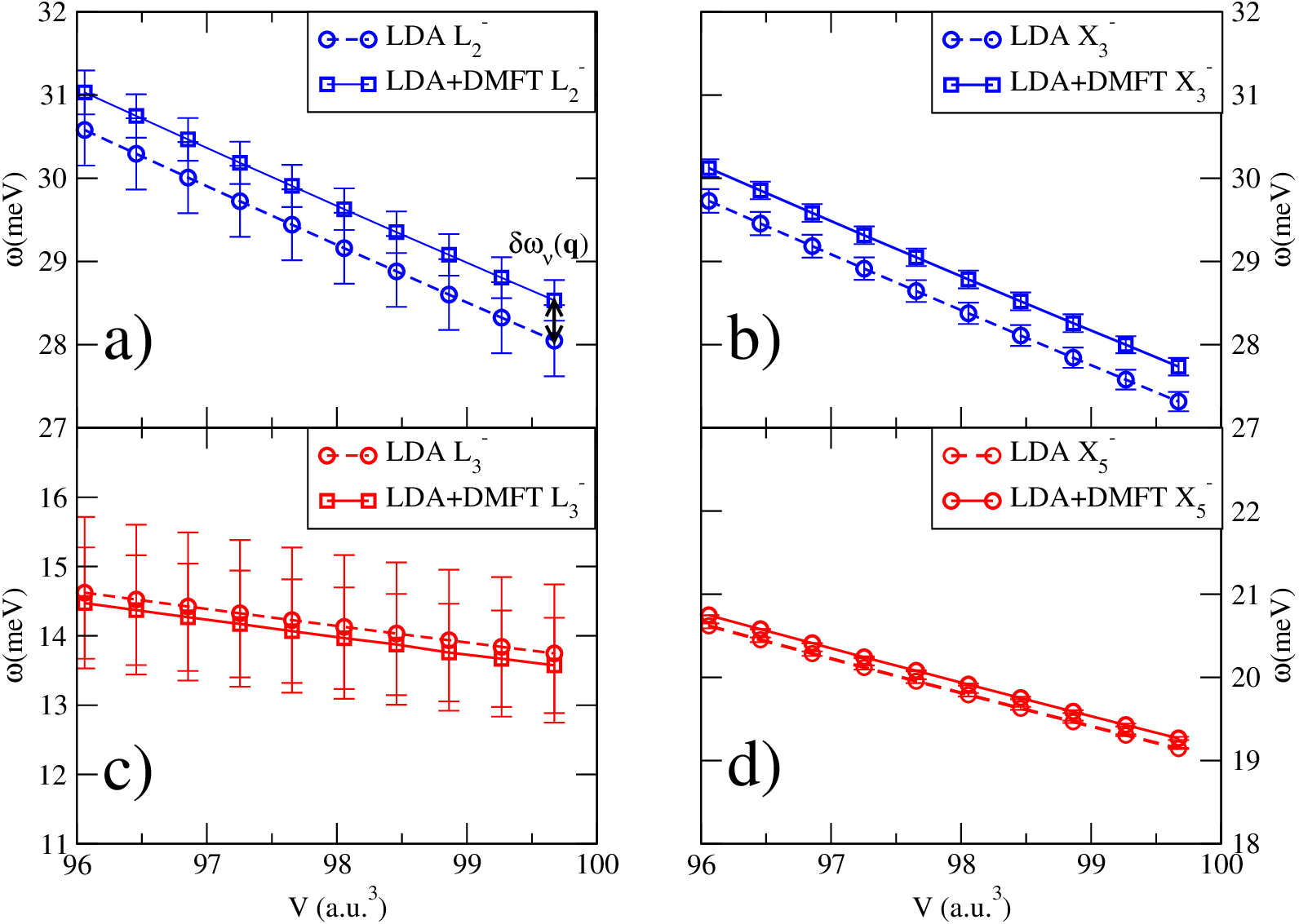} \\
        \caption{
Computed phonon frequencies at the high symmetry points $L$ (Figs.~\ref{Fig3}a,~\ref{Fig3}c) and $X$ (Figs.~\ref{Fig3}b,~\ref{Fig3}d) as a function of the volume computed in LDA (disc symbol) and LDA+DMFT (square symbol) with $U=0.75\,$eV and $J=0.3\,$eV.
The longitudinal modes are $L_2^-$ and $X_3^-$  (Fig.~\ref{Fig3}a,~\ref{Fig3}b blue color) and the transversal modes are $L_3^-$ and $X_5^-$ (Figs.~\ref{Fig3}c,~\ref{Fig3}d red color).
        }
        \label{Fig3}
\end{figure}

In Fig.~\ref{Fig3} we present the phonon frequencies at the zone boundaries
as a function of the unit-cell volume $V$ in LDA and LDA+DMFT.
Red and blue decorated lines correspond to transversal and longitudinal modes, respectively.
Upon lattice expansion (increasing volume) phonon modes are seen to become softer.
Note that in the studied range the phonon frequencies decrease almost linearly
with the volume, and within a good approximation the linear curves
are shifted rigidly by an amount $\delta \omega_\nu({\bf q}=X,L)=\omega_\nu(\mathbf{q})^{\mathrm{DMFT}}-\omega_\nu(\mathbf{q})^{\mathrm{LDA}}$ (see also Tab.~\ref{tab:gruen}).
These results indicate that the correlation effects lead to a stiffening of the lattice.
As seen in Fig.~\ref{Fig3}, the frequency shifts due to volume expansion are significant in the considered volume range in comparison with the error estimates for the  $L_2^{-},X_3^{-},X_5^{-}$-modes, while this is not the case for the $L_3^{-}$-mode.
%
%
In Tab.~\ref{tab:gruen} we present the values of the $\gamma_\nu^G({\mathbf q})$ for the modes $X_5^-,L_3^-,X_3^-$, and $L_2^-$ in LDA and LDA+DMFT.
The results show that $\gamma_\nu^G({\mathbf q},\nu)$
is larger for the longitudinal modes $X_3^-$ and $L_2^-$ than for the transversal modes $L_3^{-}$ and $X_5^{-}$.
The overall effect of electronic correlations is the significant reduction of $\gamma_\nu^G({\mathbf q},\nu)$ for the longitudinal modes,
while for the transversal modes $\gamma_\nu^G({\mathbf q},\nu)$ remains almost unchanged ($X_5^{-}$-mode) or even increases ($L_3^{-}$-mode).

\subsection{Phonon dispersion at different temperatures}
\label{sec:Tdependence}
In the previous sections we discussed
LDA+DMFT results for phonon spectra calculated at $316\,$K.
For the wave vector ${\mathbf q}=\frac{2\pi}{a}[0.375,0.375,0]$, which is close to the nesting vector $\mathbf{q}_\text{nest}$, the phonon modes were found to become stiffer with increasing Coulomb interaction strength.

In this section we present
results of LDA+DMFT calculations performed
in a wide temperature range for
interaction parameters $U=1.00\,$eV and $J=0.30\,$eV.
We used the harmonic approximation and, consequently,
the renormalization of phonon
frequencies is only due to the
electronic correlations at
finite temperatures while the
anharmonic (phonon-phonon) terms
are ignored.

Within the LDA it is easy to compute the phonon spectra at finite temperatures since this requires only a multiplication with the Fermi function when only the electronic temperature is considered, as is the case in our investigation. By contrast, for
low temperature calculations in the LDA+DMFT approach a large number of Matsubara frequencies must be included in the impurity solver to reach convergence. This increases the
computational time very significantly. The LDA+DMFT calculation of phonons were performed using both a fine and a sparse Matsubara mesh
of 32768 and 4096 frequencies, respectively. We found that for temperatures $T$
larger than about $150\,$K the difference in the phonon frequencies computed with
the finer Matsubara mesh becomes negligible in comparison with the sparse mesh.
Hence, we will use the sparse mesh in the following to keep the computational effort
at a manageable level.

\begin{figure}[h]
\includegraphics[width=\columnwidth,clip=true]{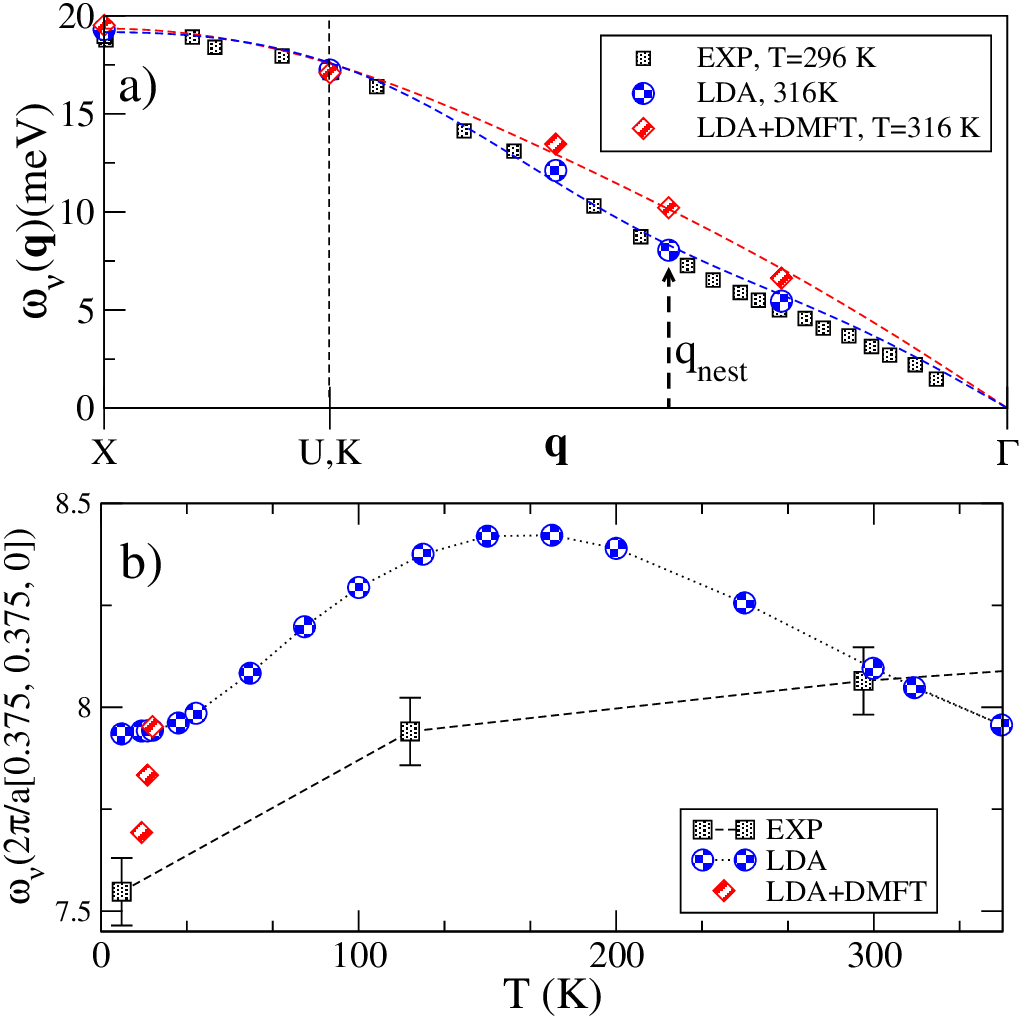} \\
\caption{a) Comparison of phonon frequencies of the
$T_1$-branch along the $\Sigma$ direction as obtained from experiment\cite{mii.bro.68} at 296~K (black
squares), LDA (blue circles), and LDA+DMFT results (red hatched diamonds).  Lines are fits and serve only as guides to the eye.
b) Comparison of the temperature dependence of the phonon frequency
$\omega_\nu({\bf q}_{nest})$ as obtained from experiment\cite{mii.bro.68,mii.bro.71} (black
squares), LDA (blue hatched circles), and LDA+DMFT (red hatched diamonds).
The interaction parameters are $U=1.00\,$eV and $J=0.30\,$eV.}
        \label{Fig4}
\end{figure}

In Fig.~\ref{Fig4}a we compare
the experimental and the computed
phonon dispersions along the
direction $\Sigma$ close to room temperature.
One clearly observes that in the vicinity of the possible Kohn anomaly at ${\bf q}_{nest}$ the DMFT results lie above the
LDA results and the experimental data \cite{mii.bro.68}.
To understand the frequency behavior near ${\bf q}_{nest}$ better we computed the temperature dependence down to low temperatures ($T = 15\,$K),
both within LDA and LDA+DMFT. The results are shown in Fig.~\ref{Fig4}b.
For temperatures $T \le 20\,$K the phonon
frequencies obtained within LDA+DMFT are found to be significantly reduced
compared to room temperature. Not only are they now lower than the LDA results
but, in fact, they approach the experimental values.
Clearly the LDA+DMFT approach performs better at lower
temperatures where quantum effects are more relevant.
The strong temperature dependence of the phonon frequencies at wave-vectors close to the nesting vector (Kohn anomaly)
in the presence of electronic correlations may be interpreted as follows. When the temperature is lowered the screening of the
atomic forces becomes more efficient, which leads to significant phonon softening. This effect cannot
be captured within LDA. The remaining discrepancies may be partly attributed to
phonon-phonon scattering not considered in the present calculations.

\section{\label{sec:conc}Conclusion}
We performed LDA and LDA+DMFT calculations and determined the phonon dispersion
$\omega_{\nu}(\mathbf{q})$, the phonon density of states, the thermodynamic Gr\"uneisen parameter $\gamma_{td}^G$, and the mode Gr\"uneisen parameters $\gamma_\nu^G(\mathbf{q})$ for zone-boundary phonon modes.
The phonon dispersion was computed using the frozen-phonon method which relies on the Born-Oppenheimer and the harmonic approximations and requires accurate total energies.

The change in the thermodynamic Gr\"uneisen parameter, $\gamma_{td}^G$,
as a function of the strength of the Coulomb $U$ and exchange $J$ interaction parameters,
demonstrates that in the presence of electron correlations anharmonic effects on the level of the QHA
become less important.

We also studied the phonon dispersion at finite temperatures in the vicinity of the possible Kohn anomaly.
In LDA+DMFT low temperature
calculations require
considerable computational effort.
We find that at  $T \sim 20$~K  the phonon frequency obtained within LDA+DMFT
has strongly decreased relative to the value at room temperature and approaches the experimental result.
We interpret this softening as being a consequence of the correlation-induced reduction of the restoring force of the ionic displacement in the harmonic approximation.

\section*{Acknowledgement}
We are grateful to Ferdi Aryasetiawan, Milos Radonji\'c and Weiwei Sun for stimulating discussions.
WHA thanks Levente Vitos for the kind hospitality at KTH Stockholm.
I.D.M. acknowledges support by the appointment to the JRG program at the APCTP through the Science and Technology Promotion Fund and Lottery Fund of the Korean Government, as well as support by the Korean Local Governments, Gyeongsangbuk-do Province and Pohang City.
Part of the  computations were performed on resources provided by the Swedish
National Infrastructure for Computing (SNIC) at Beskow.
Financial support offered by the Augsburg Center for Innovative Technologies,
and by the Deutsche Forschungsgemeinschaft (through TRR 80/F6) is gratefully acknowledged.
IL is grateful for the financial support provided by the Russian Science Foundation (project Nr. 18-12-00492).

\appendix

\section{Symmetry Adapted Dynamical Matrix}
\label{app:symmetry}

In order to study lattice vibrations in the harmonic approximation the potential energy surface is expanded in second order in the ionic displacement
\begin{eqnarray}
U_{p}(\{\mathbf{R}\})&=&E(\{\mathbf{R}^0\})\\
&=&\frac{1}{2} \sum_{\mathbf{R}\mathbf{R}'\alpha_1\alpha_2} u_{\alpha_1}(\mathbf{R})C_{\alpha_1\alpha_2}(\mathbf{R}-\mathbf{R}')u_{\alpha_2}(\mathbf{R})\nonumber
\label{eq:uharm}
\end{eqnarray}
where $u_\alpha(\mathbf{R})=R_{\alpha}-R^0_{\alpha}$ are the deviations of the ionic positions from their equilibrium positions, and $\alpha_1,\alpha_2=x,y,z$ denote the components of a vector in the 3D Cartesian space, $U_{p}(\{\mathbf{R}_n\})$ is the BO energy surface, and $C_{\alpha_1\alpha_2}(\mathbf{R}-\mathbf{R}')$ is the matrix of interatomic force constants in the Cartesian representation. We introduce the dynamical matrix
\begin{eqnarray}
D_{\alpha_1 \alpha_2 }(\mathbf{q})=\frac{1}{M} \sum_{\mathbf{R}} C_{\alpha_1 \alpha_2}(\mathbf{R}) e^{i\mathbf{q}\cdot \mathbf{R}},
\end{eqnarray}
were $M$ is the ionic mass.
The Hamiltonian of the effective phonon system $H_p=-\sum_{{\mathbf{R}}}\frac{1}{2M}\nabla^2_{{\mathbf{R}}}+U_{p}(\{\mathbf{R}\})$ is quadratic in the displacement and anharmonic terms are neglected.
The normal modes of the phonon system are determined by solving the following equation of motion:
\begin{eqnarray}
\omega^2 e_{\alpha_1}(\mathbf{q})&=&\sum_{{\alpha_1}} D_{\alpha_1\alpha_2} ({\mathbf q}) e_{\alpha_2}(\mathbf{q}),
\label{eq:eomphon}
\end{eqnarray}
where $e_{\alpha}(\mathbf{q})$ are components of the normalized vector in 3D Cartesian vector space.
Non-trivial solutions are obtained by solving the secular equation ${\rm det}\left(D_{\alpha_1\alpha_2} ({\mathbf q})-\delta_{\alpha_1\alpha_2}\omega^2\right)=0$.
The solutions are denoted as $\omega_\nu^2({\mathbf q})$ and the corresponding eigenmodes $e_{\alpha,\nu}(\mathbf{q}) $ are given by the nullspace of the matrix $D_{\alpha_1,\alpha_2}(\mathbf{q})-\omega_\nu^2({\mathbf q})\delta_{\alpha_1,\alpha_2}$:
\begin{eqnarray}
\omega_\nu^2(\mathbf{q}) e_{\alpha_1,\nu}(\mathbf{q})&=&\sum_{{\alpha_1}} D_{\alpha_1\alpha_2} ({\mathbf q}) e_{\alpha_2,\nu}(\mathbf{q}).
\label{eq:eomphon2}
\end{eqnarray}
The index $\nu=1,2,3$ enumerates the 3 linear independent phonon branches in Pd.
The solution of the secular equation can be simplified by making use of the symmetry of the system.
It turns out that the eigenvalue problem in Eq.~\ref{eq:eomphon} can be decomposed into one-dimensional blocks along the certain high symmetry lines in the irreducible Brillouin zone.
This allows us to determine the normal modes $e_{\alpha,\nu}(\mathbf{q})$ by applying group theoretical techniques without explicitly obtaining the solution of the secular equation.

\begin{figure}[t]
\includegraphics[width=\columnwidth,clip=true]{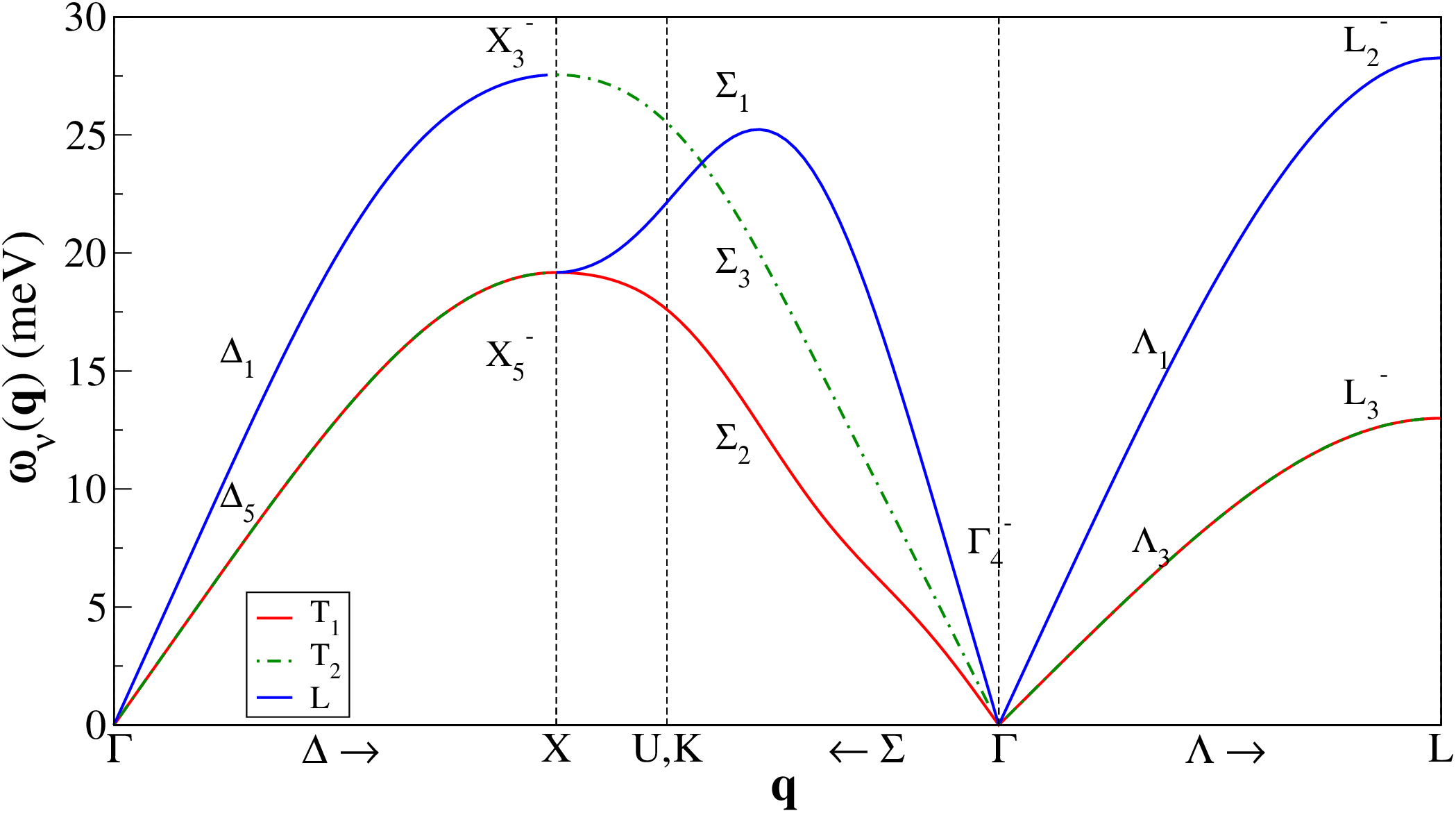}
        \caption{Phonon dispersion where the irreducible representations (irreps) are indicated. The irrep symbols follow the convention of Miller and Love~\cite{mi.67}.
        The phonon branches are also labeled by indicating the two transversal branches $T_1$, $T_2$, and the longitudinal branch $L$.
        }
        \label{Fig0}
\end{figure}

The phonon dispersion curves for a five nearest neighbor force-constant model is shown in Fig~\ref{Fig0}. The irreducible representations (irreps) $\gamma$ are indicated in the figure.
When the dynamical matrix $D_{\alpha_1,\alpha_2}(\mathbf{q})$ is written in the symmetry adapted basis then the eigenvalue problem in Eq.~\ref{eq:eomphon} decomposes into $m_\alpha$ blocks according to the generalized Wigner-Eckart theorem~\cite{lu.12,wi.93}:
 \begin{eqnarray*}
  D_{\alpha_1\alpha_2}(\mathbf{q})&\equiv&\langle \alpha_1| D(\mathbf{q})|\alpha_2\rangle  \\
  & = & \sum_{\nu\nu'}  S^{*}_{\alpha_1\nu'} S_{\nu\alpha_2} \langle \nu' | D(\bf q)|\nu\rangle\\
  & = & \sum_{\nu\nu'}  S^{*}_{\alpha_1\nu'} S_{\nu\alpha_2}  D_{\nu'\nu}(\bf q)\\
 & = & \sum_{\nu\nu'}  S^{*}_{\alpha_1\nu'} S_{\nu\alpha_2} \langle \gamma',s_\gamma,l|D(\mathbf{q})|\gamma,s_\gamma,i\rangle \\
  &=& \sum_{\nu\nu'}  S^{*}_{\alpha_1\nu'} S_{\nu\alpha_2}\delta_{\gamma',\gamma} \delta_{l,i} \langle s_\gamma'|| D(\mathbf{q})  || s_\gamma\rangle^{(i)},
 \end{eqnarray*}
here we introduced the multi-index $\nu=\gamma,s_\gamma,l$ (also the branch index in our case) which runs over all irreps $\gamma$, the row $l$ of the irrep and the the multiplicity $s_\gamma=1,\dots,m_\gamma$ of an irrep $\gamma$ in the basis $\alpha$.
The unitary transformation from the Cartesian to the symmetry adapted basis is denoted as $S_{\nu\alpha}=\langle \nu| \alpha\rangle$.
The symmetry-adapted basis is obtained employing the projecting technique of Ref.~\onlinecite{lu.12}.
Due to the highly symmetric space-group and the simple lattice basis, consisting only of a single atom per unit cell, the dynamical matrix decomposes into blocks of dimension $m_\alpha=1$.
Therefore the phonon frequencies can be computed efficiently in the frozen phonon approach since only a single symmetry mode needs to be set up in order to determine the mode frequency.
The supercell of the subgroup has to be commensurate with the wave-length ${\mathbf q}$ and ions located at  low symmetric Wyckoff positions are displaced according to the symmetry mode of interest.

The construction of the symmetry adapted basis functions is a cumbersome task since one needs to perform the projection of the basis $\alpha$ onto space group irreps for  various ${\mathbf q}$-points in the irreducible Brillouin zone.
It is useful to interpret the symmetry modes as order parameter directions of a structural second-order phase transition, and the potential energy as a function of the ionic displacement as the Landau free energy expansion~\cite{landau1968statistical,landau1937theory}.
The advantage of this approach is that the order parameters of a phase transition can be found without directly referring to the atomic positions and instead focusing on specific irreps and their invariant subspaces.
The invariant subspaces determine subgroups, the so called isotropy groups, of the high symmetric parent structure.

Stokes and Hatch
have developed a software package which allows one to obtain information about the normal
modes of oscillations in a crystal by employing the concept of isotropy subgroups~\cite{st.ha.84,ha.sto.87,st.hat.91,st.ha.88,st.95}.
By listing the number of irreps of the space-group $Fm\overline{3}m$ at a given ${\mathbf q}$-point one can determine the symmetry modes systematically.
Using one irrep at a time we project out modes which transform like basis functions of that irrep~\cite{st.hat.91,st.95}.
The supercells for the frozen-phonon calculations can be generated using the ISOTROPY Software Suite~\cite{isotropy}.
In the supercell geometries the ionic positions are shifted rigidly, where we used three displacement amplitudes of $0.02$, $0.06$, and $0.1$ atomic units.

\bibliography{sample}
\end{document}